\documentclass[a4paper,11pt]{article}
\usepackage{pos}
\newcommand{\etal}{\textit{et al.}}

\usepackage[
  per-mode=symbol-or-fraction,
]{siunitx}
\usepackage{amsmath}
\usepackage{braket}
\usepackage{pgfplots}
\pgfplotsset{compat = newest}
\usepgfplotslibrary{fillbetween}
\usepackage{subcaption}

\usepackage{graphicx}
\usepackage{grffile}

\usepackage{scrhack}
\usepackage{float}
\usepackage[section, below]{placeins}

\usepackage{subcaption}
\usepackage{mdframed}

\usepackage{tikz}
\usetikzlibrary{positioning}
\usetikzlibrary{arrows,shapes,snakes,decorations.pathmorphing, decorations.markings}
\usepackage{tikz-feynman, contour}
\tikzset{cross/.style={cross out, draw=black, fill=none, minimum size=2*(#1-\pgflinewidth),
inner sep=0pt, outer sep=0pt}, cross/.default={2pt}}
\tikzset{axis/.style={thick, ->, >=stealth'}}
\tikzset
 {managed/.unknown/.style={},
  managed/.style={name=#1,managed/#1}
 }
\tikzfeynmanset{compat=1.0.0,
every blob/.style={draw=green!40!black, pattern color=green!40!black},
}
\tikzset{/handlers/.provide style/.code={%
    \pgfkeysifdefined{\pgfkeyscurrentpath/.@cmd}{}%
        {\pgfkeys {\pgfkeyscurrentpath /.code=\pgfkeysalso {#1}}}%
}}

\tikzset{test/.style={rectangle}}
\tikzset{New/.style={rectangle, draw=cyan,text=cyan}}
\usepackage{tkz-euclide}

\renewcommand{\logo}{\relax}

\title{\begin{picture}(0,0)(0,0)%
     \put(360,100){\makebox(0,0)[l]{\textnormal{\normalsize KEK-CP-0400}}}%
\end{picture}
$B_s \to K\ell\nu$ form factors from lattice QCD with domain-wall heavy quarks}

\manuallySeparateAuthors
\author*[a]{Protick Mohanta}
\author[a,b,c]{Takashi Kaneko}
\author[a,b]{and Shoji Hashimoto}
\author{\\ (JLQCD Collaboration)}

\affiliation[a]{High Energy Accelerator Research Organization (KEK),
  Tsukuba 305-0801, Japan}

\affiliation[b]{SOKENDAI (The Graduate University for Advanced Studies), Tsukuba 305-0801, Japan}

\affiliation[c]{Kobayashi-Maskawa Institute for the Origin of Particles and the Universe,\\
Nagoya University, Aichi 464-8602, Japan}

\emailAdd{protick@post.kek.jp}

\abstract{We report on our on-going study of the $B_s \to K\ell\nu$ decay in $N_f=2+1$
lattice QCD. We employ fully relativistic setup in which 
the M\"obius domain-wall action is used for all quark flavors.
The lattice cutoff is $a^{-1} \sim 2.5$ GeV,
where we take bottom quark masses up to $m_Q\!<\!0.7a^{-1}$ in order
to control discretization errors.
We present preliminary results for the relevant
form factors extracted from correlator ratios by inspecting their ground state saturation.}

\FullConference{The 40th International Symposium on Lattice Field Theory (Lattice 2023)\\
July 31st - August 4th, 2023\\
Fermi National Accelerator Laboratory\\}

\begin{document}
\maketitle

\section{Introduction and Motivation}

The exclusive $B\!\to\!\pi\ell\nu$ decay has been conventionally used to
determine the Cabibbo-Kobayashi-Maskawa matrix element $|V_{ub}|$. 
However, about 2\,$\sigma$ tension with the inclusive analysis~\cite{pdg_rev}
suggests that theoretical and/or experimental uncertainties have not yet been
fully understood.
In our previous study~\cite{brian},
largest theoretical uncertainties came from the statistics and chiral extrapolation.
The $B_s\!=\!K\ell\nu$ decay provides an alternative determination of $|V_{ub}|$
with advantages that
i) the statistical fluctuation of relevant correlation functions
is suppressed as discussed in Refs.~\cite{lepage, parisi}
and ii) the chiral extrapolation may be better controlled 
without the valence pions.
The relevant form factors have been, therefore, calculated
by several groups~\cite{hpqcd,rbcukqcd_old,fermilabmilc,alpha,rbcukqcd_new}
using bottom quark actions based on effective theories. However,
recent LHCb measurement of $B_s^0\!\to\!K^-\mu^+\nu_\mu$
and their analysis together with $B_s^0\!\to\!D_s^-\mu^+\nu_\mu$
obtained inconsistent results for $|V_{ub}|/|V_{cb}|$
depending on the form factor inputs,
namely those from lattice QCD and light-cone sum rule~\cite{lhcb}.
Since a tension among the from factors from previous lattice QCD studies
has also been suggested in Ref.~\cite{fermilabmilc},
independent calculations are highly welcome.
In this article,
we report on our on-going study employing fully relativistic setup.

\section{Gauge Ensemble and Simulation Parameters}

We simulate $N_f = 2 +1$ QCD using the M\"{o}bius domain-wall quark action~\cite{moebius}.
Details about our ensembles can be found in Ref.~\cite{brian}.
This article reports preliminary results obtained on a $32^3\!\times\!64\!\times\!12$ lattice
at the lattice cutoff of $a^{-1}\!=\!2.453(44)$~GeV fixed from the Yang-Mills gradient flow.
The degenerate up and down quark mass $m_{ud}$ corresponds to
an unphysically large pion mass $M_\pi\!=\!499(1)$~MeV, whereas
the strange quark mass $m_s$ is close to the physical value
leading to the kaon mass $M_K\!=\!618(1)$~MeV.
The statistics are 5,000 Hybrid Monte Carlo trajectories.
The $B_s\!\to\!K$ form factors are calculated in fully relativistic setup
with bottom quark masses $m_Q\!=\!m_c, 1.25m_c, 1.25^2m_c < 0.7a^{-1}$,
where $m_c$ represents the physical charm quark mass fixed
from the spin-averaged charmonium mass $(M_{\eta_c}+3M_{J/\Psi})/4$,
so that discretization errors remain under control.

\section{Correlation Functions and Form Factors}

In order to extract the form factors,
we make use of the following three and two-point functions
\begin{eqnarray}
  C_{3,\mu}^{K \to B_s}(t,T,\vec{q})
  &=&
  \frac{1}{N_{t_0}}
  \sum_{\vec{x},\vec{y},\vec{z},t_0}
  \langle
    \mathcal{O}_{B_s}(\vec{x},T+t_0)
    V_{bl}^\mu(\vec{y},t+t_0)
    \mathcal{O}_{K}(\vec{z},t_0)^\dagger
  \rangle
  \exp{[-i \vec{q}(\vec{y}-\vec{z})]},
  \label{eqn:3pt:KtoBs}
  \\
  C_{3,\mu}^{B_s \to K}(t,T)
  &=&
  \frac{1}{N_{t_0}}
  \sum_{\vec{x},\vec{y},\vec{z},t_0}
  \langle
    \mathcal{O}_{K}(\vec{x},T+t_0)
    V_{lb}^\mu(\vec{y},t+t_0)
    \mathcal{O}_{B_s}(\vec{z},t_0)^\dagger
  \rangle,
  \label{eqn:3pt:BstoK}
  \\
  C_{3,\mu}^{K \to K}(t,T)
  &=&
  \frac{1}{N_{t_0}}
  \sum_{\vec{x},\vec{y},\vec{z},t_0}
  \langle
    \mathcal{O}_{K}(\vec{x},T+t_0)
    V_{ll}^\mu(\vec{y},t+t_0)
    \mathcal{O}_{K}(\vec{z},t_0)^\dagger
  \rangle, 
\end{eqnarray}
\begin{eqnarray}
  C_{3,\mu}^{B_s \to B_s}(t,T)
  &=&
  \frac{1}{N_{t_0}}
  \sum_{\vec{x},\vec{y},\vec{z},t_0}
  \langle
    \mathcal{O}_{B_s}(\vec{x},T+t_0)
    V_{bb}^\mu(\vec{y},t+t_0)
    \mathcal{O}_{B_s}(\vec{z},t_0)^\dagger
  \rangle,
  \label{eqn:3pt:BstoBs}
  \\
  C_2^K(t,\vec{p}_K)
  &=&
  \frac{1}{N_{t_0}}
  \sum_{\vec{x},\vec{z},t_0}
  \langle
    \mathcal{O}_{K}(\vec{x},t+t_0)
    \mathcal{O}_{K}(\vec{z},t_0)^\dagger
  \rangle
  \exp{[-i\vec{p}_{K}(\vec{x}-\vec{z})]},
  \label{eqn:2pt:K}
  \\
  C_2^{B_s}(t)
  &=&
  \frac{1}{N_{t_0}}
  \sum_{\vec{x},\vec{z},t_0}
  \langle
    \mathcal{O}_{B_s}(\vec{x},t+t_0)
    \mathcal{O}_{B_s}(\vec{z},t_0)^\dagger
  \rangle,
  \label{eqn:2pt:Bs}
\end{eqnarray}
where $V_{bl}^\mu(x)=\bar{b}(x)\gamma^\mu l(x)$, $V_{lb}^\mu(x)=\bar{l}(x)\gamma^\mu b(x)$,
$V_{ll}^\mu(x)=\bar{l}(x)\gamma^\mu l(x)$ and $V_{bb}^\mu(x)=\bar{b}(x)\gamma^\mu b(x)$.
We apply Gaussian smearing to the interpolating fields
$\mathcal{O}_{B_s}$ and $\mathcal{O}_{K}$
to enhance their overlap with the ground state.
The three-point functions are measured with five different values
of the source-sink separation $T\!=\!12, 16, 20, 24$ and 28
in order to study their ground state saturation.
The statistical accuracy of both three- and two-point functions
is improved by averaging over the source point $(\vec{z},t_0)$.
A momentum-projected volume source with $Z_2$ noise is used 
for the average over the spatial source point $\vec{z}$.
We repeat the measurement for four values of the source timeslice
$t_0\!=\!0,16,32$ and $48$ for the average over $t_0$.
These three- and two-point functions can be expressed as, for instance,
\begin{eqnarray}
  C_{3,4}^{K \to B_s}(t,T,\vec{q})
  &=&
  \sum_{n,m}
  A_n^{B_s}\,\left(A_m^{K}\right)^\ast D_{4,nm}^{K \to B_s}
  \exp{[-E_n^{B_s}(T-t)]} \exp{[-E_m^{K} t]},
  \label{eqn:3pt}
  \\
  C_2^K(t,\vec{p}_K)
  &=&
  \sum_{n}
  A_n^K\,\left(A_n^{K}\right)^\ast  \big(\exp{[-E_{K}^n t]} + \exp{[-M_{K}^n (T_l-t)]}\big),
  \label{eqn:2pt}
\end{eqnarray}
where
$A_n^{B_s}= \frac{\langle 0\vert \mathcal{O}_{B_s} \vert B_s^n\rangle}{\sqrt{2 E_n^{B_s}}}$
and $A_n^{K}= \frac{\langle 0\vert \mathcal{O}_{K}\vert K^n\rangle}{\sqrt{2 E_n^{K}}}$
are the overlaps of the interpolating fields with the physical states,
and $D_{4,nm}^{K \to B_s}= \frac{\langle B_s^n\vert V^4_{bl}\vert K^m
\rangle}{2\sqrt{ E_n^{B_s} E_m^K} } $ involves the matrix element of interest.

The $B_s$ meson is at rest throughout our measurement.
Then parallel and perpendicular form factors in the parametrization
\begin{eqnarray}
  \langle K(p_K) | V^\mu | \langle B(p_{B_s}) \rangle
  & = &
  \sqrt{2 M_{B_s}} \left\{ v^\mu f_\parallel(E_K) + p_{K,\perp}^\mu f_\perp(E_K) \right\}
\end{eqnarray}
are given by the following simple expressions
\begin{equation}
  f_{\parallel}(E_{K})=\frac{\langle K|V^{0}|B_{s}\rangle}{\sqrt{2M_{B_{s}}}},
  \hspace{5mm}
  f_{\perp}(E_{K})=\frac{1}{p_{K}^{i}}\,\frac{\langle K|V^{i}|B_{s}\rangle}{\sqrt{2M_{B_{s}}}}.
\end{equation}
In order to study the $E_K$ dependence,
namely momentum transfer dependence, of the form factors,
we vary the kaon momentum as $|\vec{p}_K|^2\!=\!0,1,2,3$ (in units of $(2\pi/L)^2$),
and calculate $C_{3,\mu}^{K \to B_s}$ and $C_2^K(t,\vec{p_K})$
as indicated in Eqs.~(\ref{eqn:3pt:KtoBs}) and (\ref{eqn:2pt:K}).
For the latter, we used the local kaon sink,
which is useful for canceling overlap factors in a correlator ratio for non-zero recoils
(see Eq.~(\ref{eqn:r4}) below).
Other correlators (\ref{eqn:3pt:BstoK})\,--\,(\ref{eqn:3pt:BstoBs})
and (\ref{eqn:2pt:Bs}) are used only for zero recoil.

\section{Form Factors Extraction at Zero Recoil}
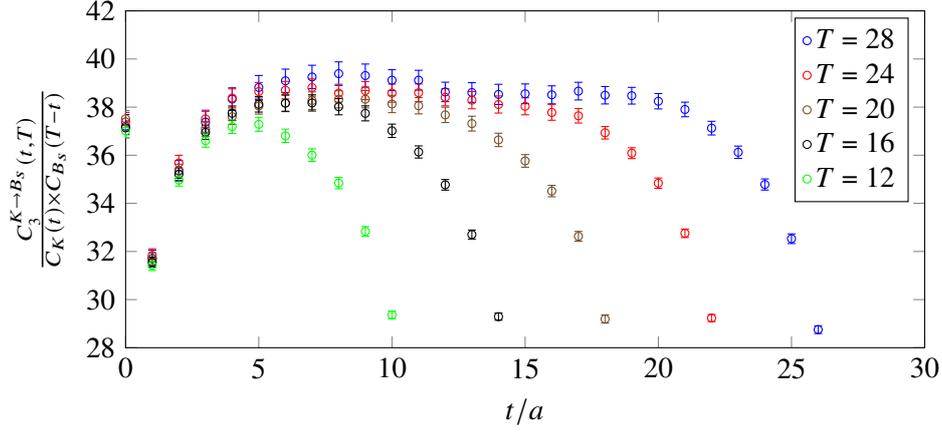
\begin{figure}[!ht]
\begin{center}
\begin{tikzpicture}
\begin{axis}[
    xmin = 0, xmax = 30,
    ymin = 28, ymax = 42,
    xtick distance = 5.0,
    ytick distance = 2.0,
    width = 0.8\textwidth,
    height = 0.4\textwidth,
    xlabel = {$t/a$},
    ylabel = {$C_3^{K \rightarrow B_s}(t,T)
    \over C_{K}(t)\times C_{B_s}(T-t)$}
    ]
\addplot+ [only marks, mark = o, mark size = 1.5pt]
        plot [error bars/.cd, y dir = both, y explicit]
	table [x ={x}, y ={y}, y error ={dy}] {T_28.dat};
\addplot+ [only marks, mark = o, mark size = 1.5pt]
        plot [error bars/.cd, y dir = both, y explicit]
	table [x = x, y = y, y error = dy] {T_24.dat};
\addplot+ [only marks, mark = o, mark size = 1.5pt]
        plot [error bars/.cd, y dir = both, y explicit]
	table [x = x, y = y, y error = dy] {T_20.dat};
\addplot+ [only marks, mark = o, mark size = 1.5pt]
        plot [error bars/.cd, y dir = both, y explicit]
	table [x = x, y = y, y error = dy] {T_16.dat};
\addplot+ [green, only marks, mark = o, mark size = 1.5pt]
        plot [error bars/.cd, y dir = both, y explicit]
	table [x = x, y = y, y error = dy] {T_12.dat};

\legend{
      $T$ = 28,
      $T$ = 24,
      $T$ = 20,
      $T$ = 16,
      $T$ = 12
       }
\end{axis}
\end{tikzpicture}
\caption{Correlator ratio $R_{3p2p}$ as a function of the temporal location of the vector
current $t$. We plot data with $m_Q =  m_c$.}
\label{r3p2p}
\end{center}
\end{figure}

We test two correlator ratios to extract the form factor at zero recoil.
Figure~\ref{r3p2p} shows the following ratio of three- to two-point functions
\begin{equation} R_{3p2p} =
 \frac{C_{3,4}^{K \to B_s}(t,T)}{C_2^{K}(t)\, C_2^{B_s}(T-t)} \rightarrow \frac{D_{4,00}^{K
     \to B_s} }{ A_0^K \, A_0^{B_s^\ast} }
\label{eqn:r3p2p} 
\end{equation}
as a function of the temporal location of the vector current.
We observe reasonable ground state saturation particularly
around the temporal mid-point $t\!\sim\!T/2$. 
To estimate the ground state contribution in the rightmost side of Eq.~(\ref{eqn:r3p2p}),
we carry out a simultaneous fit of the three- and two-point function to the expressions 
(\ref{eqn:3pt})\,--\,(\ref{eqn:2pt})
including the ground ($n,m\!=\!0$) and first excited ($n,m\!=\!1$) states.
This ratio, however, needs
the renormalization constant of the heavy-light current $Z_{V_{bl}}$,
which we estimate as $Z_{V_{bl}}\!=\!\sqrt{Z_{V_{ll}} Z_{V_{bb}}}$.
For the determination of $Z_{V_{bb}}$ we use the vector current conservation
$Z_{V_{bb}} C_{3,4}^{B_s \to B_s}(t,T)/C_2^{B_s}(T)\!=\!1$
and $Z_{V_{ll}}$ is determined in our study \cite{tomii}.

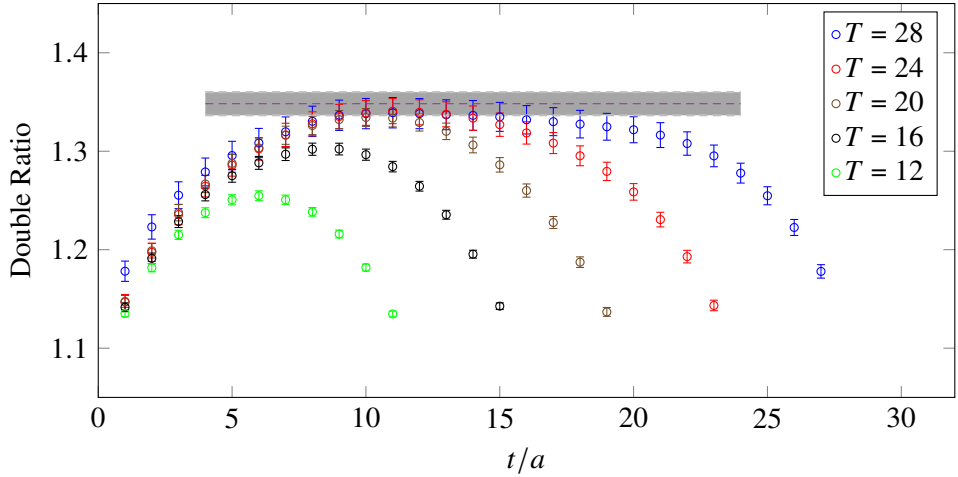
\begin{figure}
\begin{center}
 \begin{tikzpicture}
\begin{axis}[
    xmin = 0, xmax = 32,
    ymin = 1.05, ymax = 1.45,
    xtick distance = 5.0,
    ytick distance = 0.1,
    width = 0.85\textwidth,
    height = 0.45\textwidth,
    xlabel = {$t/a$},
    ylabel = {Double Ratio}
    ]
\addplot+ [only marks, mark = o, mark size = 1.5pt]
        plot [error bars/.cd, y dir = both, y explicit]
	table [x ={x}, y ={y}, y error ={dy}] {cntr_err_5T_28};
\addplot+ [only marks, mark = o, mark size = 1.5pt]
        plot [error bars/.cd, y dir = both, y explicit]
	table [x = x, y = y, y error = dy] {cntr_err_5T_24};
\addplot+ [only marks, mark = o, mark size = 1.5pt]
        plot [error bars/.cd, y dir = both, y explicit]
	table [x = x, y = y, y error = dy] {cntr_err_5T_20};
\addplot+ [only marks, mark = o, mark size = 1.5pt]
        plot [error bars/.cd, y dir = both, y explicit]
	table [x = x, y = y, y error = dy] {cntr_err_5T_16};
\addplot+ [green, only marks, mark = o, mark size = 1.5pt]
        plot [error bars/.cd, y dir = both, y explicit]
	table [x = x, y = y, y error = dy] {cntr_err_5T_12};
\addplot+ [domain=4:24, smooth, thick, no markers, name path=A, gray!50] {1.336525};
\addplot+ [domain=4:24, smooth, thick, no markers, name path=B, gray!50] {1.359992};
\addplot+ [domain=4:24, smooth, no markers,  violet!80] {1.348259};
\addplot+ [gray!70] fill between[of=A and B];

\legend{
      $T$ = 28,
      $T$ = 24,
      $T$ = 20,
      $T$ = 16,
      $T$ = 12
       }
\end{axis}
\end{tikzpicture}
 \caption{Double ratio at $m_Q = m_c$.
   Symbols are simulation results for five different
   values of the source-sink separation $T$, whereas the gray band shows coefficient
   $C_{00}$ in Eq.~(\ref{drat_fit}) representing the ground state contribution.}
\label{dblerat}
\end{center}
\end{figure}

We also test the double ratio of three-point functions 
\begin{equation}
\textmd{Double Ratio} = \frac{C_{3,4}^{K \to B_s}(t,T)\, C_{3,4}^{B_s \to K}(t,T)}
{C_{3,4}^{K \to K}(t,T)\, C_{3,4}^{B_s \to B_s}(t,T)}\rightarrow \frac{D_{4,00}^{K \to B_s}\,
D_{4,00}^{B_s \to K} }{D_{4,00}^{K \to K}\, D_{4,00}^{B_s \to B_s}} \; ,
\end{equation}
where the renormalization and overlap factors get canceled~\cite{shoji}.
We carry out a simultaneous fit of this ratio to the form 
\begin{eqnarray}
 \frac{C_{3,4}^{K \to B_s}(t,T)\, C_{3,4}^{B_s \to K}(t,T)}{C_{3,4}^{K \to K}(t,T)\,
C_{3,4}^{B_s \to B_s}(t,T)} &=&  C_{00} \Big(1 + A^{\prime}\Big[ \exp{[-\Delta E_{K} t]} +
\exp{[-\Delta E_{K}(T-t)]} \Big] \nonumber\\
&& + ~~ B^{\prime}\Big[\exp{[-\Delta E_{B_s} t]} + \exp{[-\Delta E_{B_s}(T-t)]} \Big]\Big)
\label{drat_fit}
\end{eqnarray}
and two-point functions to Eq.~(\ref{eqn:2pt})
to estimate the coefficients $C_{00}$, $A^\prime$ $B^\prime$
and energy differences $\Delta E_{B_s}$ and $\Delta E_K$.
Simulation data of the ratio and its ground state contribution $C_{00}$
are plotted in Fig.~\ref{dblerat},
where we again observe reasonable ground state saturation around the mid-point
$t\!\sim\!T/2$.

\begin{table}[h]
\begin{center}
\begin{tabular}{|c|c|c|c|}
 \hline
 $m_Q$ & $m_c$ & $1.25m_c$ & $1.25^2 m_c$  \\
 \hline
 $R_{3p2p}A_0^K \, A_0^{B_s} \sqrt{Z_{V_{bl}}}$ & 1.154(12) & 1.177(12) & 1.209(13)\\
 $\sqrt{\textmd{Double Ratio}}$ & 1.1546(10) & 1.1802(16) & 1.2061(25)\\
 \hline
\end{tabular}
\caption{Comparison of $R_{3p2p}$ and Double Ratio}
\label{Tab: Comparison}
\end{center}
\end{table}

The ground state contributions of
$R_{3p2p}$ and the double ratio are related as
\begin{equation}
R_{3p2p}A_0^K \, A_0^{B_s} \sqrt{Z_{V_{bl}}}  =  \sqrt{\textmd{Double Ratio}}.
\end{equation}
Good numerical consistency shown in Table~\ref{Tab: Comparison} suggests that
the ground state contribution is reliably extracted for both of the two ratios.
The same table also shows that the result with $R_{3p2p}$ has much larger uncertainty,
which mainly comes from the measurement of $Z_{V_{ll}}$.
Hence we have used double ratio in our analysis.

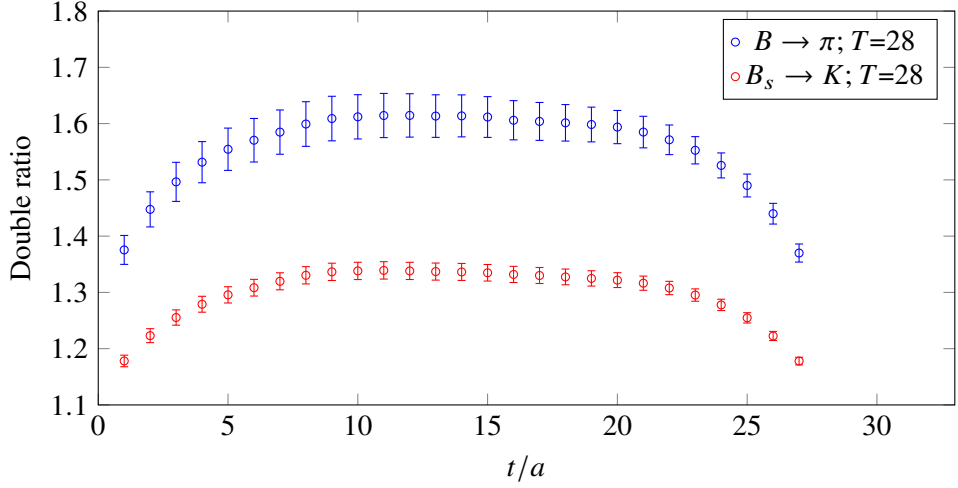
\begin{figure}
\begin{center}
\begin{tikzpicture}
\begin{axis}[
    xmin = 0, xmax = 33,
    ymin = 1.1, ymax = 1.8,
    xtick distance = 5.0,
    ytick distance = 0.1,
    width = 0.85\textwidth,
    height = 0.45\textwidth,
    title = ,
    xlabel = {$t/a$},
    ylabel = {Double ratio}
    ]
\addplot+ [only marks, mark = o, mark size = 1.5pt]
        plot [error bars/.cd, y dir = both, y explicit]
	table [x ={x}, y ={y}, y error ={dy}] {cntr_err_5T_28_piB};
\addplot+ [only marks, mark = o, mark size = 1.5pt]
        plot [error bars/.cd, y dir = both, y explicit]
	table [x = x, y = y, y error = dy] {cntr_err_5T_28};

\legend{
       $B \rightarrow \pi$; $T$=28,
       $B_s \rightarrow K$; $T$=28
       }
\end{axis}
\end{tikzpicture}
\caption{Double ratios of $B \to \pi$ and $B_s \to K$ at $m_Q = m_c$.
 We plot data for the source sink separation $T=28$.}
\label{vs_b_to_pi}
\end{center}
\end{figure}

Our accuracy of the form factor at zero recoil is typically 0.5\,\%.
In accordance with the discussions in Refs.~\cite{lepage, parisi},
this is significantly better than that for $B\!\to\!\pi\ell\nu$
as shown in Fig.~\ref{vs_b_to_pi}.

\section{Form Factor Extraction for Non-zero Recoil}

In order to extract {$f_\parallel(E_K)$} and and for $f_\perp(E_K)$ at non-zero recoils,
we consider the following ratios respectively
\begin{eqnarray}
R_4(\vec{p}_K) &=& \frac{C_{3,4}^{K \rightarrow B_s}(t,T,\vec{p}_{B_s}=\vec{0},\vec{p}_K)}
{C_{3,4}^{K \rightarrow B_s}(t,T,\vec{p}_{B_s}=\vec{0},\vec{p}_K=\vec{0})}\times
\frac{C_2^K(t,\vec{p}_K=\vec{0})}{C_2^K(t,\vec{p}_K)},
 \label{eqn:r4}
\\
R_i(\vec{p}_K) &=& \frac{C_{3,i}^{K \rightarrow B_s}(t,T,\vec{p}_{B_s}=\vec{0},\vec{p}_K)}
{C_{3,4}^{K \rightarrow B_s}(t,T,\vec{p}_{B_s}=\vec{0},\vec{p}_K)},
\end{eqnarray}
where three- and two-point functions are averaged
over appropriate momentum configurations to improve their statistical accuracy.
We then carry out a combined fit of these ratios to the following forms
\begin{eqnarray}
R_4(\vec{p}_K) &=&  C_{44} \Big(1 + A \exp{[-\Delta E_{B_s}(\vec{p}_{B_s}=\vec{0})(T-t)]} +
 \nonumber \\
 && B \exp{[-\Delta E_{K}(\vec{p}_K=\vec{0})t]} + F \exp{[-\Delta E_{K}(\vec{p}_K)t]}  \Big),
 \\
 R_i(\vec{p}_K) 
&=&  C_{4i} \Big(1 + G \exp{[-\Delta E_{B_s}(\vec{p}_{B_s}=\vec{0})(T-t)]} +
 H \exp{[-\Delta E_{K}(\vec{p}_K)t]}  \Big),
\end{eqnarray}
and two-point functions, $C_2^K(t,\vec{p}_K)$ and $C_2^{B_s}(t)$ to Eq.~(\ref{eqn:2pt}).
The coefficients $A$, $B$, $F$, $G$ and $H$ are originated
from the excited state contributions.
The overall factors, $C_{44}$ and $C_{4i}$, encode the ground state contribution
and can be used to determine the form factors as 
\begin{equation}
  f_{\parallel}(E_{K})= C_{44}\sqrt{2M_K C_{00}}, \hspace{5mm}
  f_{\perp}(E_{K})= \frac{C_{4i}C_{44}
\sqrt{2M_K C_{00}}}{p^i_K}.
\end{equation}

\begin{center}
 \begin{figure}[!ht]
    \begin{subfigure}[t]{.52\textwidth}
\includegraphics[width=\textwidth, height=0.3\textheight]{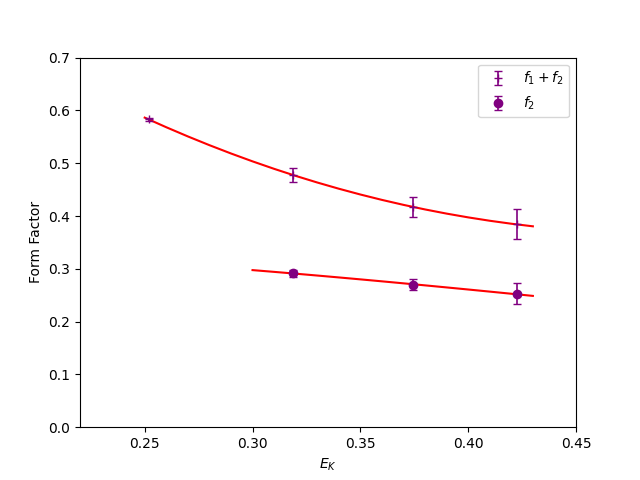}
\caption{$m_Q$=$m_c$}
\label{form_fact1}
    \end{subfigure}%
    \begin{subfigure}[t]{.52\textwidth}
\includegraphics[width=\textwidth, height=0.3\textheight]{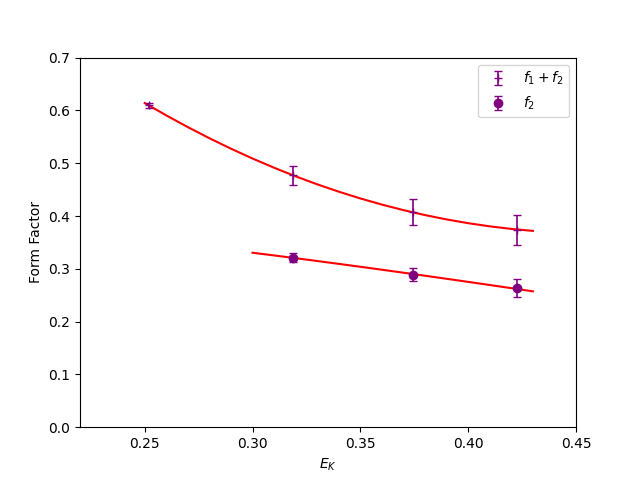}
\caption{$m_Q$=$1.25^2 m_c$}
\label{form_fact2}
    \end{subfigure}%
    \caption{Form factors $f_1+f_2$ (pluses) and $f_2$ (circles)
      as a function of kaon energy $E_K$.
      The left and right panels show results at $m_Q\!=\!m_c$ and $1.25^2m_c$,
      respectively.
      The red lines are fit curves polynomial in $E_K$.
    }
    \end{figure}
\end{center}

As in our study of $B\!\to\!\pi\ell\nu$~\cite{brian},
we convert $f_\parallel$ and $f_\perp$ to
HQET motivated definition of form factors $f_1$ and $f_2$ as
 \begin{equation}
   f_1(E_K) + f_2(E_K) = {f_\parallel(E_K)\over \sqrt{2}},
   \hspace{5mm}
  f_2(E_K) = {E_K\; f_\perp(E_K) \over \sqrt{2}}.
 \end{equation}
In Figs.~\ref{form_fact1} and \ref{form_fact2},
we plot these form factors as a function of $E_K$.
The typical statistical accuracy is
\,7\,\% for $f_1+f_2$
and \,8\,\% for $f_2$,
which is better than the $B\!\to\!\pi\ell\nu$ case.

\section{Summary and Outlook}
We report on JLQCD's on-going study of the $B_s \to K\ell\nu$ decay
with the M\"obius domain-wall bottom quarks.
We present preliminary results at $a^{-1}\!\sim\!2.5$\,GeV and $M_\pi\!\sim\!500$~MeV
with bottom quark masses taken up to $0.7a^{-1}$ in order to control discretization effects.
The ground state saturation has been carefully confirmed
by simulating five values of the source-sink separation
and testing two correlator ratios to extract the form factors.
The statistical accuracy is improved by averaging relevant correlators
over four source timeslices,
and turns out to be better than our previous study of $B\!\to\!\pi\ell\nu$
using the same gauge ensemble.
The extension of this study
to pion masses down to $M_\pi \gtrsim 230$ MeV
and two larger lattice cutoffs $a^{-1} \lesssim 4.5$ GeV is on-going.

\section*{Acknowledgement}
We thank members of the JLQCD collaboration for fruitful discussions.
This work used computational resources of supercomputer
Fugaku provided by the RIKEN Center for Computational Science
(Project IDs: hp220140 and hp230122),
SX-Aurora TSUBASA at the High Energy Accelerator Research
Organization (KEK) under its Particle, Nuclear and Astrophysics Simulation Program
(Project ID: 2023-004),
the FUJITSU Supercomputer PRIMEHPC FX1000 and FUJITSU Server PRIMERGY GX2570
(Wisteria/BEDECK-01) at the Information Technology Center, The University of Tokyo
through MCRP (Project ID: Wo22i049).
This work is supported in part by JSPS KAKENHI Grant Numbers JP21H01085
and JP22H00138 .

\end{document}